\documentclass[11pt]{article}
\usepackage[sc]{mathpazo}
\usepackage{fullpage}
\usepackage[authoryear,sectionbib,sort]{natbib}
\usepackage{adjustbox}
\usepackage{multirow}
\usepackage{amsmath,amssymb,amsfonts}
\usepackage[utf8]{inputenc}
\usepackage{lineno}
\usepackage{titlesec}
\usepackage{graphicx}
\usepackage{subcaption}
\usepackage{booktabs,multirow}
\usepackage{here}
\usepackage{url}
\usepackage{multirow}
\usepackage{algorithm}
\usepackage{algorithmic}

\titleformat{\section}[block]{\Large\bfseries\filcenter}{\thesection}{1em}{}
\titleformat{\subsection}[block]{\Large\itshape\filcenter}{\thesubsection}{1em}{}
\titleformat{\subsubsection}[block]{\large\itshape}{\thesubsubsection}{1em}{}
\titleformat{\paragraph}[runin]{\itshape}{\theparagraph}{1em}{}[. ]

\title{Towards species' classification of the \textit{Anastrepha pseudoparallela} group}

 \author{Gabriel R. Palma$^{1, 2, \ast}$ \and Rocío Alaiz$^{5}$ \and Alexandre S. Araújo$^{4}$
 \and Marcoandre Savaris$^{4}$ \and Roberto A. Zucchi$^{4}$\and Charles Markham$^{3}$\and
Rafael A. Moral$^{1, 2}$}

\date{}

\begin{document}

\maketitle

\noindent{} 1. Hamilton Institute, Maynooth University, Maynooth, Ireland;

\noindent{} 2. Department of Mathematics and Statistics, Maynooth University, Maynooth, Ireland;

\noindent{} 3. Department of Psychology, Maynooth University, Maynooth, Ireland;

\noindent{} 4. Department of Entomology and Acarology, University of São Paulo, Piracicaba, Brazil.

\noindent{} 5. Department of Electrical Engineering, Systems and Automation, Universidad de León.León.Spain

\noindent{} $\ast$ Corresponding author; e-mail: gabriel.palma.2022@mumail.ie

\bigskip


\bigskip

\bigskip

\textit{Manuscript type}: Research paper. 

\bigskip

\noindent{\footnotesize Prepared using the suggested \LaTeX{} template for \textit{Am.\ Nat.}}

\newpage{}

\section*{Abstract}
Identifying \textit{Anastrepha} species from the \textit{pseudoparallela} group is problematic due to morphological similarities among species and a broad geographic variation. This group comprises $31$ species of fruit flies and includes pests that affect passion fruit crops. The identification of these species utilises the morphological characteristics of the specimens' wings and aculeus tips. Considering the importance of identifying these species of this group, this paper contributes to automating the classification of the species \textit{Anastrepha chiclayae} Greene, \textit{Anastrepha consobrina} (Loew), \textit{Anastrepha curitibana} Araújo, Norrbom \& Savaris, \textit{Anastrepha curitis} Stone, and \textit{Anastrepha pseudoparallela} (Loew) using deep Learning methods and designing new features based on wing structures. Automating the classification of this group is a challenge due to the low data availability and the imbalance of classes in the dataset. We explored transfer learning solutions and proposed a new set of features based on each wing's structure. We used the dual annealing algorithm to optimise the hyperparameters of Deep Neural Networks, Random Forests, Decision Trees, and Support Vector Machines algorithms combined with autoencoders and SMOTE algorithms to account for class imbalance. We tested our approach with a dataset of 127 high-quality images belonging to a total of $5$ species and used three-fold cross validation for training, tuning and testing, encompassing six permutations of those to assess the performance of the learning algorithms. Our findings demonstrate that our novel approach, which combines feature extraction and machine learning techniques, can improve the species classification accuracy for rare \textit{Anastrepha pseudoparallela} group specimens, with the SMOTE and Random Forests algorithms leading to the average performance of $0.72$ in terms of mean of the individual accuracies considering all species. Our results are promising in classifying rare species characterised by small, imbalanced datasets.

\textit{Keywords}: Transfer learning, Machine learning, Data augmentation, Class Imbalance Feature engineering, DNN Architecture optimisation.

\newpage{}

\section{Introduction} 


The classification and identification of insects is fundamental for comprehending global biodiversity, given that this group represents an expressive percentage of the available biodiversity today~\citep{gaston1991magnitude, sankarganesh2017insect, wagner2021insect, chowdhury2023protected, vaz2023global, hailay2024systematic}. A clear picture of the global biodiversity yields a better understanding of the ecological services that new species can potentially provide and enlightens conservation and management practices~\citep{thrupp2004importance, gamfeldt2008multiple, uchida2021urban, upreti2023importance, riva2024principles}. Entomologists play a fundamental role in identifying taxa and understanding the evolutionary, ecological and functional relationships among these taxa~\citep{pryke2024routledge}. The study of biodiversity is heavily reliant upon correct classification of animal specimens. The identification can be performed in various ways, using morphometric features~\citep{el2024wing, rodrigues2024molecular, laojun2024outline}, DNA-based identification, such as the barcoding method~\citep{chua2023future, srivathsan2024ontbarcoder}, acoustic features~\citep{chen2014flying, phung2017automated, hibino2021classification, he2024enhancing, branding2024insectsound1000} and other methods~\citep{raffini2020nucleotides, hoye2021deep, van2022emerging, karbstein2024species}. In this context, there is an opportunity to apply different methods targeting insect automatic monitoring~\citep{van2022emerging, hoye2021deep, palma2023machine}.

Various researchers have analysed the efficiency of machine learning, deep learning, computer vision and a combination of these methods when applied to insect classification, especially for pest species~\citep{passias2024insect, assiri2024automated}. The results obtained from these studies have been positively affecting Integrated Pest Management (IPM) protocols by providing data-driven decision-making systems~\citep{gao2024application, moonis2024optimized, amrani2024multi}. However, challenges have been reported related to the use of these techniques to automate insect monitoring, such as the high similarities of insect species, obtaining large and high-quality annotated datasets, and the presence of imbalanced classes due to lack of homogeneity in the same groups of insects ~\citep{nawoya2024computer, pise2024imbalanced}. 

In other research areas, several authors have proposed adaptations of deep learning methods to accommodate the low data availability, such as changes in Convolutional Neural Networks (CNN) architectures by including regularisation techniques, such as drop out and $L_1$ re\-gu\-la\-ri\-za\-tion~\citep{brigato2021close, koppe2021deep}. Transfer learning is commonly reported as an alternative solution to increase the performance of deep learning models for image classification~\citep{barbero2024addressing, rachman2024enhanced}. Also, several data-augmented approaches have been implemented to increase the number of samples, such as Synthetic Minority Over-sampling Technique (SMOTE), autoencoders (AEs), variational autoencoders (VAEs), generative adversarial networks (GANs) and adaptations of deep generative modelling methods using additional layers such as convolution, max pooling and others~\citep{van2024deep, mumuni2024survey}. In entomology, researchers have recently proposed the use of augmentation techniques to increase the number of samples for species of fruit fly~\citep{shen2024open, medina2024object, zhang2024enhancing}. Also, the use of transfer learning is a common practice for the classification of fruit flies~\citep{leonardo2018deep,martins2019deep,gosaye2022mobile, slim2023smart, molina2023remote}.

For imbalanced datasets, several techniques have been implemented to deal with this issue, such as SMOTE, undersampling, oversampling, and combining transfer learning and active sampling~\citep{yuan2023review, liu2023imbalanced, wongvorachan2023comparison, rezvani2023broad}. Specifically, rotation and change in the background are commonly reported in the literature as alternatives for data augmentation. In entomology, researchers have implemented these techniques for different taxa, belonging to Diptera, Coleoptera, Hymenoptera, and Lepidoptera~\citep{bjerge2023hierarchical, pise2024imbalanced, doan2023large}. Moreover, the use of generative deep learning methods, such as autoencoders, variational autoencoders, and other variations, has been commonly reported in the literature as an alternative to generating new data \citep{cabrera2021investigating, klasen2022image, borowiec2022deep, nitin2023developing, phong2024classification}. These techniques have shown promising results, and more researchers have started implementing them in their automation frameworks~\citep {al2024innovative, khan2024review}.

One example of the combination of imbalanced and low data availability challenges is the \textit{Anastrepha pseudoparallela} group (Diptera: Tephritidae)~\citep{araujo2024new}. Identifying \textit{Anastrepha} species from this group is problematic due to morphological similarities among species and a broad geographic variation~\citep{araujo2024new}. This group comprises $31$ species of fruit flies and includes pests that affect passion fruit crops~\citep{norrbom1999phylogeny, malavasi2000moscas}. Some species of this group are easily encountered, and others are rare in the field, generating a class imbalance problem for this group. 

Therefore, we propose a new framework to identify species of this group, specifically, \textit{Anastrepha chiclayae} Greene, \textit{Anastrepha consobrina} (Loew), \textit{Anastrepha curitibana} Araújo, Norrbom \& Savaris, \textit{Anastrepha curitis} Stone, and \textit{Anastrepha pseudoparallela} (Loew) due to the agronomic and ecological importance of this group combined with the challenge posted at the quantitative methods for automating the monitoring of these species~\citep{araujo2024new}. This paper aims to propose a new framework combining the wing's morphological features collection with machine learning algorithms and data augmentation techniques to automate the identification of species of the \textit{Anastrepha pseudoparallela} group. We explored the use of different learning algorithms, preprocessing, feature extraction methods and augmentation techniques to provide a novel approach to identifying species of this group. 

The rest of this paper is organised as follows. In Section~\ref{featureExtraction}, we introduce the proposed approach for extracting and collecting features from the images. In Section~\ref{DataAugmentationMethods}, we present the main methods used for data augmentation. In section~\ref{DLMLmethods}, we introduce the learning algorithms used to classify the five species of the \textit{pseudoparallela} group and the selected approach for hyperparameter tuning. Finally, in Section~\ref{ExperimentalResult}, \ref{discussion}, and \ref{conclusions}, we, respectively, present the experimental results, discuss our findings and present our overall conclusions from this paper.

\section{Methods}
\begin{figure}[!ht]
    \centering
    \begin{center}
 		\includegraphics[width=0.7\textwidth]{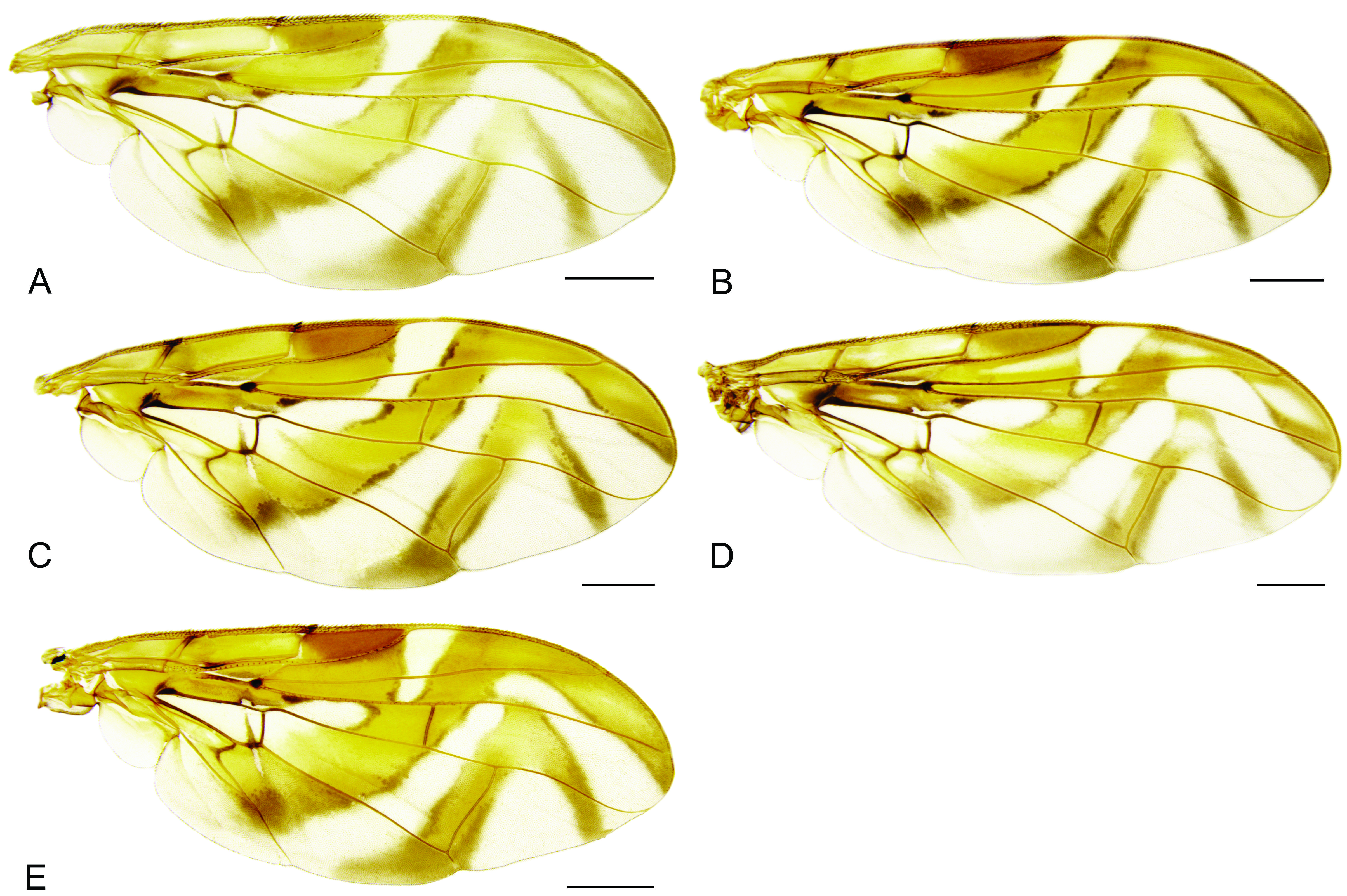}
 	\end{center}
	
 	\caption{Wings of (A) \textit{Anastrepha chiclayae}; (B) \textit{Anastrepha consobrina}; (C) \textit{Anastrepha curitibana}; (D) \textit{Anastrepha curitis}; (E) \textit{Anastrepha pseudoparallela}. Scale bars = 1.00 mm}
    \label{wings_illustraion}
\end{figure}

\subsection{Image processing and feature extraction}
\label{featureExtraction}
For image acquisition, the right wing of $127$ females from five species of the \textit{Anastrepha pseudoparallela} group was first detached from the thorax and submerged in Celossolve (C4H10O2) for 3-5 days. Then, it was mounted on permanent slides containing Euparal® and dried for seven days in a laboratory oven at 25ºC. The wing pattern of each specimen was photographed with a Leica DFC 450 camera coupled with a Leica M205 stereomicroscopic. The final dataset is composed by $21$, $16$, $72$, $11$ and  $7$ images of respectively \textit{A. curitibana}, \textit{A. pseudoparallela}, \textit{A. chiclayae}, \textit{A. curitis}, and \textit{A. consobrina} species. 
\begin{figure}
    \centering
    \includegraphics[width=.7\linewidth]{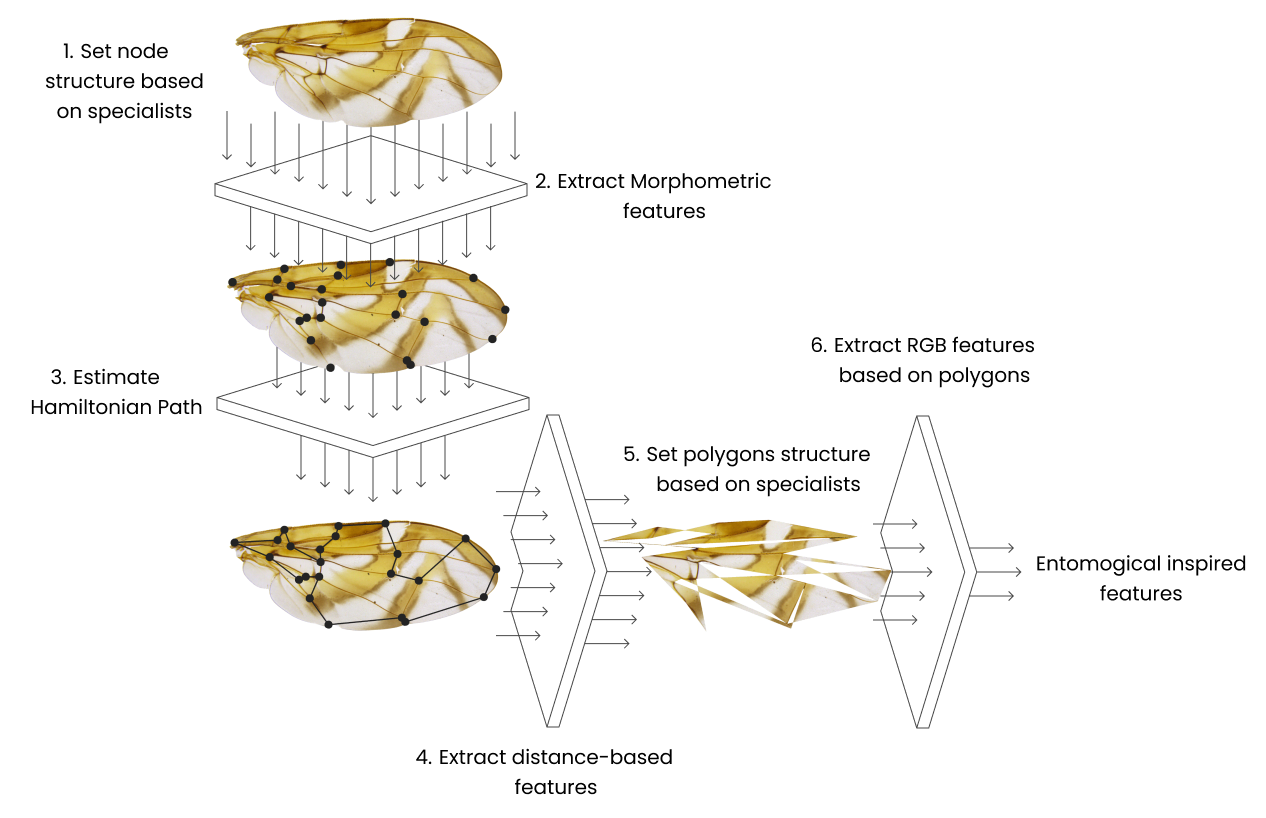}
    \caption{Diagram illustrating the proposed feature extraction using morphometric and RGB data based on the proposed approach using distance and colour-based features with, respectively, the shortest HC and polygon structures based on specialist input.}
    \label{TSPDiagram}
\end{figure}
Initially, we first removed the white background using RGB thresholding and the Canva application to remove any noise in the wing's images. Then, we preprocessed these images to create three distinct datasets composed by coloured, gray-scale, and histogram-equalizedimages~\citep{Palma2022}. We used the OpenCV package \citep{opencv2000} in most image processing techniques applied to create these datasets. Finally, we combined the Convolutional Neural Network architecture VGG16 trained with the imagenet dataset~\citep{simonyan2014} and Principal Component Analysis (PCA) for feature extraction from each wing dataset, as explained below~\citep{kaur2024vgg16, singh2024machine}.

We used this approach to avoid adverse effects on the learning algorithms' performance due to the high dimension of the feature vector produced by the VGG16 architecture. This approach allows us to explore classical machine learning and deep learning algorithms to classify the species of the \textit{Anastrepha pseudoparallela} group. The coloured wing images have dimensions of $2560 \times 1920$ pixels, which would produce a VGG16 feature vector of $2,457,600$ elements. Therefore, we first reduced the image size to $256 \times 192$ pixels, producing VGG16 feature vectors of $24,576$ elements. Finally, we applied PCA to reduce dimensionality and mitigate potential performance degradation in classical machine learning algorithms. The number of principal components retaining 95\% of the variability for the coloured, gray-scale and histogram-equalised datasets were $87$, $89$, and $87$, respectively.

We systematically collected classical morphometric features, including wing length and height measurements. Additionally, we identified features based on the structural composition of the wings. This process involved analysing the nodes and polygons constituting critical cells within the wing structure. We selected the distances between each pair of nodes and the total distance of all pairs as features. In addition, we used the polygons produced by each important cell that are taxonomic relevant to the species identification and collected the average, $2.5\%$ and $97.5\%$ percentile of RGB from all polygons designed for each image \citep{palma2023machine}.

To define the pairs of nodes, we assumed that each wing's node is a vertex of an undirected graph and identified the shortest Hamiltonian Cycle (HC) for the wing's structure. The shortest HC is a closed loop in a graph that visits each vertex exactly once and returns to the starting vertex~\citep{lozin2024hamiltonian}. Finding the shortest HC in a graph is a well-known computer science problem named the Travelling Salesman Problem (TSP)~\citep{carmesin2023hamiltonian}. In the context of our work, each node represents a key structural point of the wing. This way, we ensure that only the most relevant geometric relationships between nodes are captured while avoiding the inclusion of redundant or less informative connections. Overall, we propose using the pair-wise distances of the selected nodes belonging to the TSP's solution combined with RGB features of the selected polygons formed by regions of interest on the wings and the wing's length and height. Figure~\ref{TSPDiagram} illustrates the proposed approach to collect these features for a given shortest HC identified in a wing.

We used the Ant Colony Optimisation (ACO) algorithm to find a TSP solution for the wing's structure considering its common use in the literature to solve this problem~\citep{dorigo1997ant, stutzle1999aco, li2003dynamic}. The Ant Colony Optimisation (ACO) algorithm is a probabilistic approach inspired by the behaviour of ants seeking a path between their colony and a food source~\citep{dorigo2006ant, pedemonte2011survey, mohan2012survey, DOKEROGLU2019106040, tang2021review}. The pseudo code~\ref{ACOAlgorithm} introduces an overview of the mechaniques of this algorithm. We used $2000$ ants, $\alpha = 1$, $\beta = 2$, $\rho = 0.5$, $Q = 100$, and $200$ iterations. 

\begin{algorithm}
\caption{Wing Feature Extraction using ACO-based TSP}
\begin{algorithmic}[1]
\REQUIRE Wing image $I$
\REQUIRE ACO parameters: $N_{ants}$, $T$, $\alpha$, $\beta$, $\rho$, $Q$
\ENSURE Feature vector $F$

\STATE \textbf{Step 1: Initialize structures}
(I) Extract structural nodes $N$ from wing image $I$; (II) Extract taxonomic cell polygons $P$ from wing image $I$; (III) Initialize empty feature vector $F$

\STATE \textbf{Step 2: Extract proposed morphometric features}
(I) Calculate wing length $l$ as maximum distance between anterior-posterior nodes; (II) Calculate wing height $h$ as maximum distance between dorsal-ventral nodes; (III) Add $l,h$ to feature vector $F$

\STATE \textbf{Step 3: ACO algorithm for TSP}
(I) Initialize pheromone matrix $\tau_{ij}$ by sampling values from a uniform distribution; (II) Calculate visibility matrix $\eta_{ij} = 1/d_{ij}$ where $d_{ij}$ is distance between nodes $i,j$; (III) Initialize best solution $S_{best}$ with $L_{best} = \infty$

\FOR{$t = 1$ to $T$}
    \STATE Construct $N_{ants}$ solutions using probability $p_{ij}^k = \frac{(\tau_{ij})^\alpha(\eta_{ij})^\beta}{\sum_{l \in U_k}(\tau_{il})^\alpha(\eta_{il})^\beta}$
    \STATE Update $S_{best}$ if better solution found
    \STATE Update pheromone trails: $\tau_{ij} = (1-\rho)\tau_{ij} + \sum_k \Delta\tau_{ij}^k$
    \STATE where $\Delta\tau_{ij}^k = Q/L_k$ if edge $(i,j)$ is in ant's path
\ENDFOR

\STATE \textbf{Step 4: Extract distances from best Hamiltonian path}
(I) Calculate sequential node distances from $S_{best}$; (II) Add node distances to feature vector $F$

\STATE \textbf{Step 5: Extract colour features from polygons}
\FOR{each polygon $p$ in $P$}
    \STATE Calculate mean and percentile (2.5\%, 97.5\%) RGB values
    \STATE Add RGB statistics to feature vector $F$
\ENDFOR

\STATE \textbf{Return:} Feature vector $F$
\end{algorithmic}
\label{ACOAlgorithm}
\end{algorithm}

\subsection{Data augmentation}
\label{DataAugmentationMethods}
Considering the challenge of having solely $127$ images belonging to five classes, we addressed this problem by implementing data augmentation to increase the number of species samples with fewer images to balance the dataset. We used two approaches to perform this task: the SMOTE (Synthetic Minority Over-sampling Technique)~\citep{chawla2002smote, fernandez2018smote, khan2024review} and deep generative modelling based on an autoencoder architecture~\citep{dong2018, pratella2021} to increase the number of samples from the minority class. SMOTE is a data augmentation algorithm that generates synthetic minority class samples by interpolating between existing minority class instances in the feature space ~\citep{chawla2002smote, fernandez2018smote}. A classical autoencoder is a variation of an Artificial Neural Network (ANN) architecture, which contains an encoder, latent and decoder components, where the encoder and decoder contain ANN's layers and the same number of neurons. The latent component is an ANN's layer with fewer number of neurons, which is commonly used for encoding, dimensional reduction, and data generation~\citep{sarroff2014musical, thakkar2019autoencoder, mansouri2020laughter, yoon2022interpolation, govender2024analysing}. 

The final balanced datasets based on SMOTE and autoencoder contained $360$ and $720$ samples. The SMOTE algorithm contains fewer data points due to the algorithms' classical structure of increasing solely the minority classes~\citep{chawla2002smote, fernandez2018smote}, and we increased the number of samples of the minority classes to equal the majority class. We used the same approach for the autoencoder algorithm, and considering its flexibility~\citep{jeong2022autoencoder}, we evenly included additional samples for all classes.

\begin{figure}
    \centering
    \begin{center}
 		\includegraphics[width=0.7\textwidth]{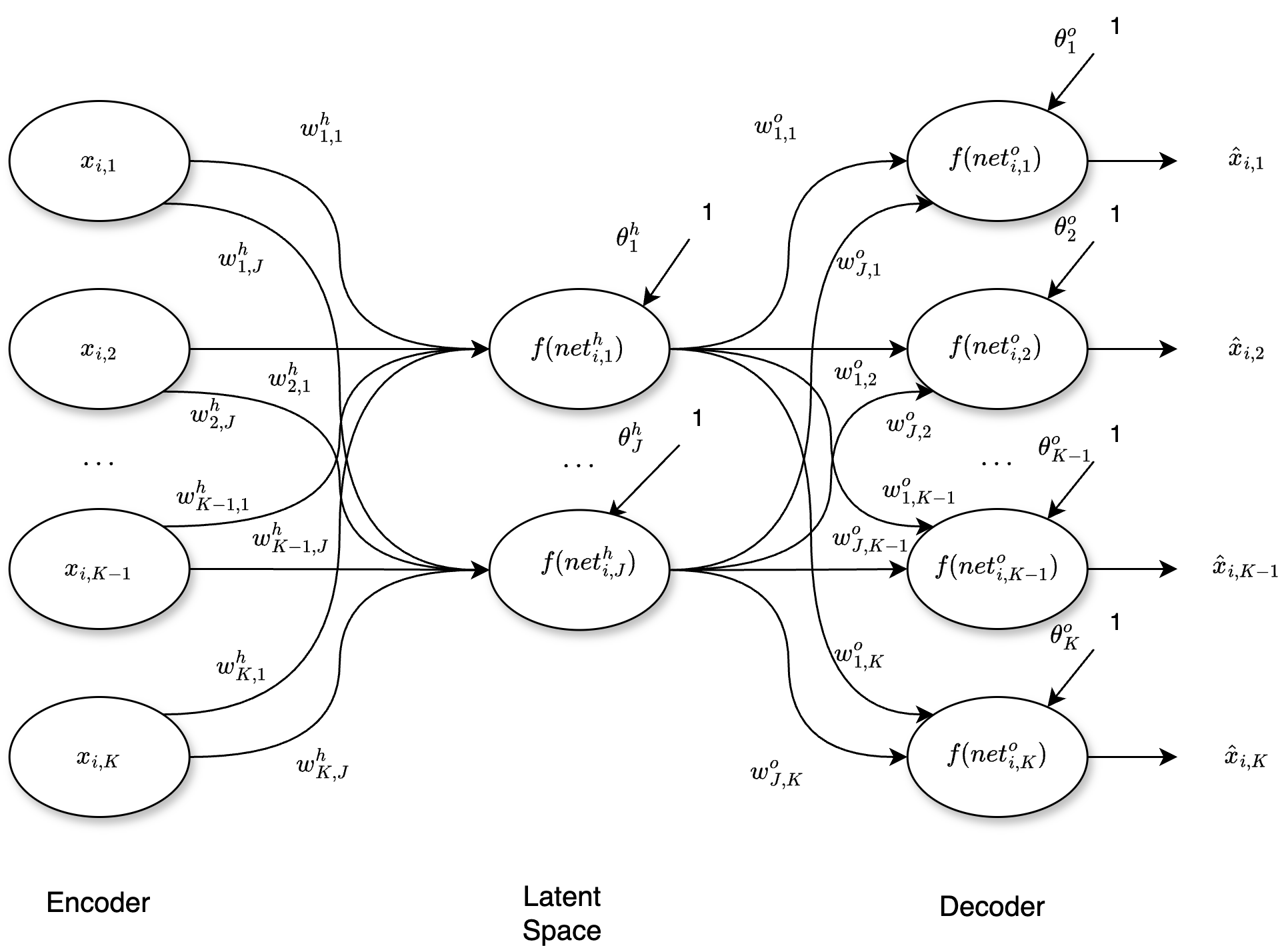}
 	\end{center}
	
 	\caption{Diagram of a simple autoencoder containing one layer with $J$ and $K$ neurons for the latent space and encoder/decoder components.}
    \label{autoencoder_diagram}
\end{figure}

Figure~\ref{autoencoder_diagram} presents a standard autoencoder architecture containing one layer for the encoder, decoder and latent components, where $w_{k, j}^{h}$ and $w_{k, j}^{o}$ are the weights, $\theta^{h}_{j}$ and $\theta^{o}_{1}$ are the biases for respectively, the encoder and decoder components. $net^{h}_{i, j}$ and $net^{o}_{i, K}$ represents linear combinations and $f$ is an activation function. We used two layers for the encoder with $n$ and $30$ neurons, where $n$ corresponds to the number of input features of a presented dataset. For the decoder, we also used two layers; the first performs batch normalisation operation, and the second contains $n$ neurons. We used a \textit{Leaky Rectified Linear Unit} (LeakyReLU) activation function, and we included a custom loss function defined as the combination of normalised Euclidean distance and the correlation coefficient between the true and predicted values~\citep{deng2017sar, wang2018distance, wu2020remove, hu2023side}. We selected the following loss function after testing a few options and evaluating the performance of the learning algorithms:

\begin{equation}
    \mathcal{L} = \frac{||\mathbf{y}_{\text{true}} - \mathbf{y}_{\text{pred}}||_2}{\sqrt{\sum \left(\max(\mathbf{y}_{\text{true}}) - \min(\mathbf{y}_{\text{true}})\right)^2} + \epsilon} - \rho(\mathbf{y}_{\text{true}}, \mathbf{y}_{\text{pred}}),
\end{equation}

where $\rho(.)$ is the algebraic correlation between the true, $\mathbf{y}_{\text{true}}$, and predicted, $\mathbf{y}_{\text{pred}}$, output vectors, respectively. $||\mathbf{y}_{\text{true}} -\mathbf{y}_{\text{pred}}||_2$ represents the Euclidean distance and $\epsilon$ is a small constant to prevent division by zero. This loss function minimises the distance, thereby reducing the reconstruction error while simultaneously maximising the correlation between the true and predicted values to preserve the inherent relationships within the data. Finally, we used the \textit{Adadelta} algorithm~\citep{zhang2023dive} to update the parameters of the autoencoder structure. We varied the autoencoder structure to explore the effects of a multi-headed attention mechanism\citep{vaswani2017attention} on the data augmentation process and, consequently, the learning algorithm's performance.

Finally, to visualise the geometric and neighbourhood structures of the augmented datasets used for classification, the Pairwise Controlled Manifold Approximation Projection (PaCMAP) method~\citep{wang2021} was used, considering its capability to preserve both local and global structures \citep{wang2021, palma2023machine}.

\subsection{Validation approach and learning algorithms}
\label{DLMLmethods}
Overall, considering the pre-processing techniques applied to the images, augmentation techniques and the feature extraction procedures, we applied different learning algorithms to 12 datasets: 

\begin{itemize}
    \item Three datasets containing VGG16-PCA features from coloured, gray-scale, and histogram equalised images with augmented data based on the autoencoder algorithm;
    \item Three datasets containing VGG16-PCA features from coloured, gray-scale, and histogram equalised images with augmented data based on the SMOTE algorithm;
    \item Three datasets containing VGG16-PCA features from coloured, gray-scale, and histogram equalised images with no data augmentation;
    \item Three datasets containing the proposed entomologically-inspired features from coloured, gray-scale, and histogram equalised images without data augmentation, as well as augmented data based on the SMOTE and the autoencoder algorithms;
\end{itemize}

For the validation procedure, we split each dataset into three folds by dividing the number of samples evenly for each fold. The first fold was used for training, the second for fine-tuning and the third for obtaining the test performance to observe the generalisation potential of each algorithm. We used six permutations of these folds and calculated the mean of individual accuracies for each permutation. We implemented four learning algorithms for classifying the \textit{pseudoparallela} species, Random Forests (RF), Support Vector Machines (SVM) with polynomial kernel, Decision trees (DT) and Deep Neural Networks (DNNs). We used the Scikit-learn~\citep{scikitlearn} and TensorFlow~\citep{tensorflow2015} packages. 

During the optimisation of the hyperparameters of each algorithm, we use the dual annealing algorithm \citep{tsallis1996}, which is a stochastic algorithm used for finding the global minimum of a given function. Briefly, the objective function used to optimise the parameters of each learning algorithm computes the negative value of the average of the individual accuracies per class obtained by the learning algorithms using a test dataset. Then, the Generalised Simulated Annealing method uses $200$ visits on the objective function to find the optimum combination of parameters. Table~\ref{AlgorithmsAttributesTable} shows the hyperparameters optimised by the dual annealing algorithm, the boundaries used for each hyperparameter, and the learning algorithm's nature. Additionally, we optimised the image size used for the learning algorithms. We used Python for image processing, machine learning and data augmentation methods and R for data visualisation. All code and datasets are available at \url{https://github.com/GabrielRPalma/AnastrephaPsedoparallelaClassification} to allow full reproducibility of this work.

\begin{table}[!ht]
	\caption{Attributes of the learning algorithms used for classifying species of the \texttt{pseudoparallela} group including the type of hyperparameter and boundaries used in the dual annealing algorithm during the optimisation process.}
	\label{AlgorithmsAttributesTable}
	\centering
	\begin{adjustbox}{max width=\textwidth}
	\begin{tabular}{ccp{5cm}p{5cm}}
		\hline
		& & \multicolumn{2}{c}{Attributes}
		
		\\\cmidrule{2-4}
		Learning algorithms &  \multicolumn{1}{c}{Hyperparameters} & \multicolumn{1}{c}{Description} & \multicolumn{1}{c}{Boundaries}
		\\
		\hline
		\\[-0.1cm]		

         \textit{Decision Trees (DT)} & $j$, $s$, $c$& $j$ is the maximum depth of the created trees, $c$ is the minimum sample leafs and $c$ is the minimum sample split & $j=\{2, \ldots, 50\}$, $s=\{2, \ldots, 50\}$ and $c=\{2, \ldots, 50\}$\\[0.05cm]
  
		\textit{Deep Neural Network (DNN)}& $l$, $u$, $d$ and $\alpha$& $l$ is the number of layers, $u$ is the total number of neurons of the neural network, $d$ indicates dropout is implemented in the and $\alpha$ is the negative slope used for the \textit{Leaky Rectified Linear Unit} (LeakyReLU) activation function in the DNN architecture& $l = \{1, \ldots, 7\}$, $d = \{0, 1\}$, $u = \{10, \ldots, 50\}$ and $\alpha = \{0, \ldots, 1\}$\\[0.05cm]
  
		\textit{Support Vector Machines (SVM)} & $p$& The polynomial degree of the applied kernel used in the method& $p = \{1, \ldots, 20 \}$
		\\[0.1cm]
		\textit{Random Forest (RF)} & $b$, $j$, $s$& $b$ is an indicator to apply bootstrapping, $j$ is the maximum depth of the created trees and $s$ is the minimum sample split & $b = \{0, 1\}$, $j = \{2, \ldots, 50\}$ and $s = \{2, \ldots, 50\}$ 		
		 \\[0.1cm]
		 \\[0.1cm]
		\hline
	\end{tabular}
	\end{adjustbox}
\end{table}

\section{Experimental results}
\label{ExperimentalResult}
\begin{figure}
    \centering
    \includegraphics[width=1\linewidth]{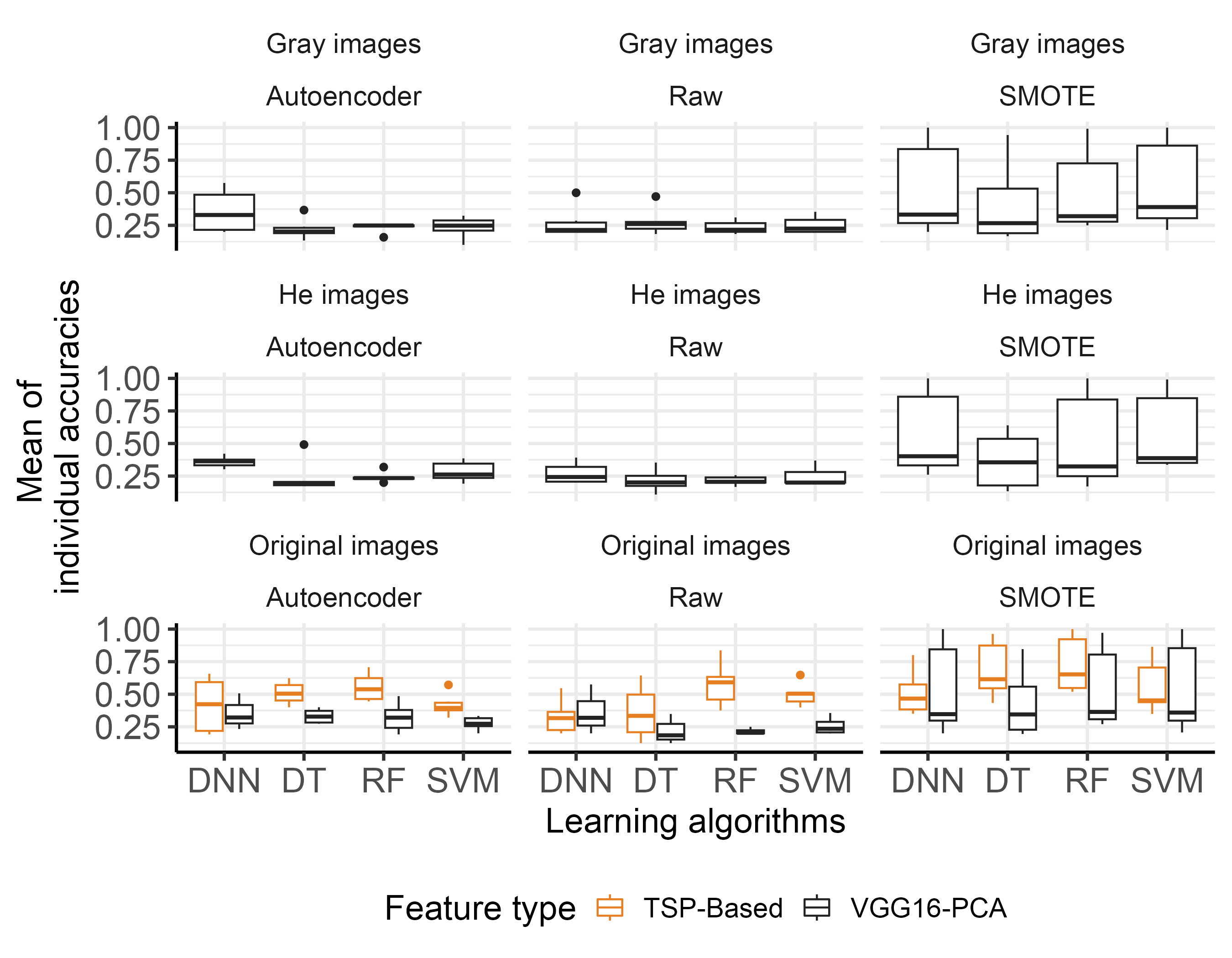}
    \caption{Mean of individual accuracies of four machine learning methods applied to the classification of \textit{A. consobrina}, \textit{A. curitis}, \textit{A. chiclayae} and \textit{A. curitibana} based on images of their wings using. All measures reported in this table were obtained in the test set.}
    \label{AlgorithmsPerformance}
\end{figure}

Figure~\ref{AlgorithmsPerformance} presents the performance obtained by the Deep Neural Networks, Support Vector Machines, Decision Trees, and Random Forests algorithms to classify species of the species \textit{Anastrepha pseudoparallela} group based on the proposed and VGG16-PCA features. Overall, considering all machine learning and data augmentation algorithms, we obtained an average and standard deviation (sd) of the mean of individual accuracies for the Travel Salesman Problem (TSP)-based and VGG16-PCA features of, respectively, $0.51~(0.19)$ and $0.35~(0.22)$. Also, considering the pre processing the images, we obtained for the coloured, gray scale and histogram equalised images an average performance of, respectively, $0.44~(0.22)$, $0.34~(0.23)$, and $0.34~(0.22)$. Moreover, we obtained an average and standard deviation (sd) of the mean of individual accuracies for SMOTE, autoencoder, and raw datasets of, respectively, $0.53~(0.29)$, $0.33~(0.13)$, and $0.30~(0.14)$. Random forests, Deep Neural Networks, Decision Trees, and Support Vector Machines algorithms obtained an average performance of $0.40~(0.25)$, $0.40~(0.22)$, $0.35~(0.21)$, and $0.39~(0.23)$.

For the classification of \textit{A. chiclayae} considering all machine learning and data augmentation algorithms, we obtained an average of the mean of individual accuracies for the TSP-based and VGG16-PCA features of, respectively, $0.78~(0.24)$ and $0.75~(0.29)$. Also, considering the pre processing the images, we obtained for the coloured, gray scale and histogram equalised images an average performance of, respectively, $0.77~(0.26)$, $0.74~(0.32)$, and $0.75~(0.27)$. Moreover, we obtained an average of the mean of individual accuracies for SMOTE, autoencoder, and raw datasets of, respectively, $0.72~(0.22)$, $0.70~(0.31)$, and $0.84~(0.21)$. In addition, Random Forests, Deep Neural Networks, decision trees, and Support Vector Machines algorithms obtained an average performance of $0.81~(0.23)$, $0.87~(0.19)$, $0.56~(0.30)$, and $0.78~(0.27)$.

For the classification of \textit{A. consobrina} considering all machine learning and data augmentation algorithms, we obtained accuracies for the TSP-based and VGG16-PCA features of, respectively, $0.32~(0.41)$ and $0.28~(0.36)$. Also, considering the pre processing the images, we obtained for the coloured, gray scale and histogram equalised images an average performance of, respectively, $0.32~(0.38)$, $0.27~0.36)$, and $0.27~(0.36)$. Moreover, we obtained accuracies for SMOTE, autoencoder, and raw datasets of, respectively, $0.52~(0.41)$, $0.23~(0.32)$, and $0.13~(0.26)$. In addition, Random Forests, deep neural network, decision tree and Support Vector Machines algorithms obtained performances of respectively $0.296~(0.38)$, $0.31~(0.39)$, $0.28~(0.36)$, and $0.29~(0.37)$

For the classification of \textit{A. curitis} considering all machine learning and data augmentation algorithms, we obtained accuracies for the TSP-based and VGG16-PCA features of, respectively, $0.70~(0.36)$ and $0.24~(0.34)$. Also, considering the pre processing the images, we obtained for the coloured, gray scale and histogram equalised images an average performance of, respectively, $0.48~(0.42)$, $0.25~(0.34)$, and $0.23~(0.33)$. Moreover, we obtained accuracies for SMOTE, autoencoder, and raw datasets of, respectively, $0.67 (0.40)$, $0.58 (0.37)$, $0.39 (0.38)$, and $0.47 (0.41)$. In addition, Random Forests, deep neural network, Decision Trees, and Support Vector Machines algorithms obtained an average performance of $0.36~(0.41)$, $0.34~(0.41)$, $0.38~(0.39)$, and $0.35~(0.39)$.

\begin{figure}
    \centering
    \includegraphics[width=.8\linewidth]{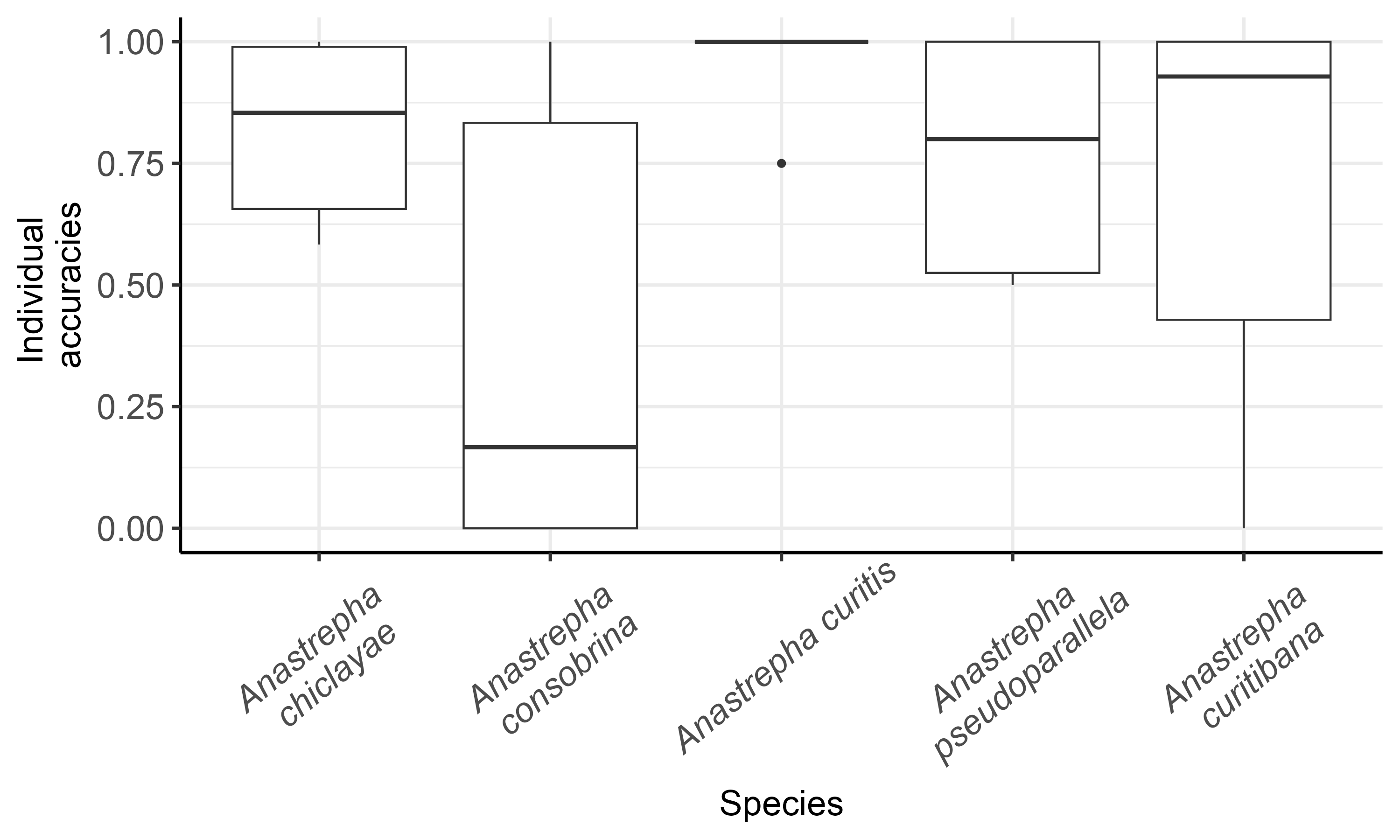}
    \caption{Random Forests' individual accuracies of each species of the \textit{Anastrepha} group based on the features collected from the coloured images of their wings using the SMOTE algorithm for data augmentation. All measures reported in this table were obtained in the test set.}
    \label{RFPerformance}
\end{figure}

For the classification of \textit{A. pseudoparallela} considering all machine learning and data augmentation algorithms, we obtained accuracies for the TSP-based and VGG16-PCA features of, respectively, $0.43~(0.36)$ and $0.20~(0.34)$. Also, considering the pre processing the images, we obtained for the coloured, gray scale and histogram equalised images an average performance of, respectively, $0.32~(0.36)$, $0.20~(0.34)$, and $0.20~(0.35)$. Moreover, we obtained an average of the mean of individual accuracies for SMOTE, autoencoder, and raw datasets of, respectively, $0.47~(0.42)$, $0.212~(0.30)$, and $0.10~(0.23)$. In addition, Random Forests, deep neural network, Decision Trees, and Support Vector Machine algorithms obtained an average performance of $0.25~(0.35)$, $0.19~(0.34)$, $0.30~(0.36)$, and $0.29~(0.38)$.

For the classification of \textit{A. curitibana} considering all machine learning and data augmentation algorithms, we obtained accuracies for the TSP-based and VGG16-PCA features of, respectively, $0.32 (0.36)$ and $0.25~(0.32)$. Also, considering the pre processing the images, we obtained for the coloured, gray scale and histogram equalised images an average performance of, respectively, $0.30~(0.35)$, $0.24~(0.34)$, and $0.25~(0.29)$. Moreover, we obtained accuracies for SMOTE, autoencoder, and raw datasets of, respectively, $0.41~(0.40)$, $0.22~(0.27)$, and $0.19~(0.28)$. In addition, Random Forests, deep neural network, Decision Trees, and Support Vector Machines algorithms obtained performances of $0.30~(0.35)$, $0.29~(0.38)$, $0.25~(0.30)$, and $0.24~(0.30)$.

Figure~\ref{RFPerformance} shows the individual accuracies per species of the studied group, where the Random Forests algorithm was employed using the coloured images for feature collection and the SMOTE algorithm. This combination presents the highest average performance we could obtain for the classification of this group, yielding an average performance of individual accuracies of $0.72~(0.22)$. The average individual accuracies performance for \textit{A. chiclayae}, \textit{A. consobrina}, \textit{A. curitis}, \textit{A. pseudoparallela}, and \textit{A. curitibana} were, respectively, $0.82~(0.19)$, $0.39~(0.49)$, 
$0.96~(0.10)$, $0.77~0.26$, and $0.69~(0.44)$

\begin{figure}
    \centering
    \includegraphics[width=1\linewidth]{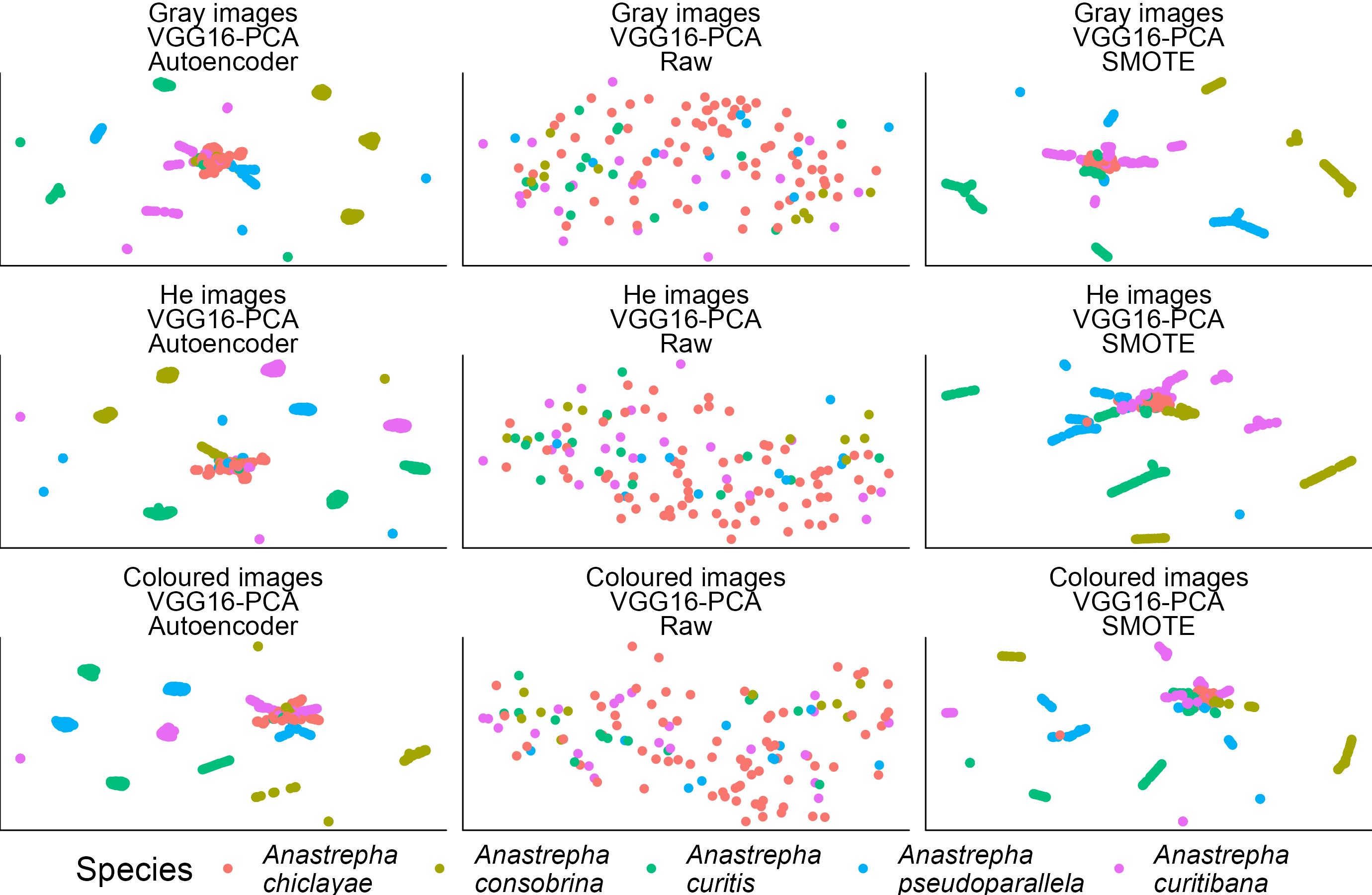}
    \caption{PaCMAP visualisation applied to the augmented VGG16-PCA features. The colour represents each species of the \textit{pseudoparallela} group presented in the dataset.}
    \label{VGG16AgumentedFeaturePaCMAP}
\end{figure}

\begin{figure}
    \centering
    \includegraphics[width=1\linewidth]{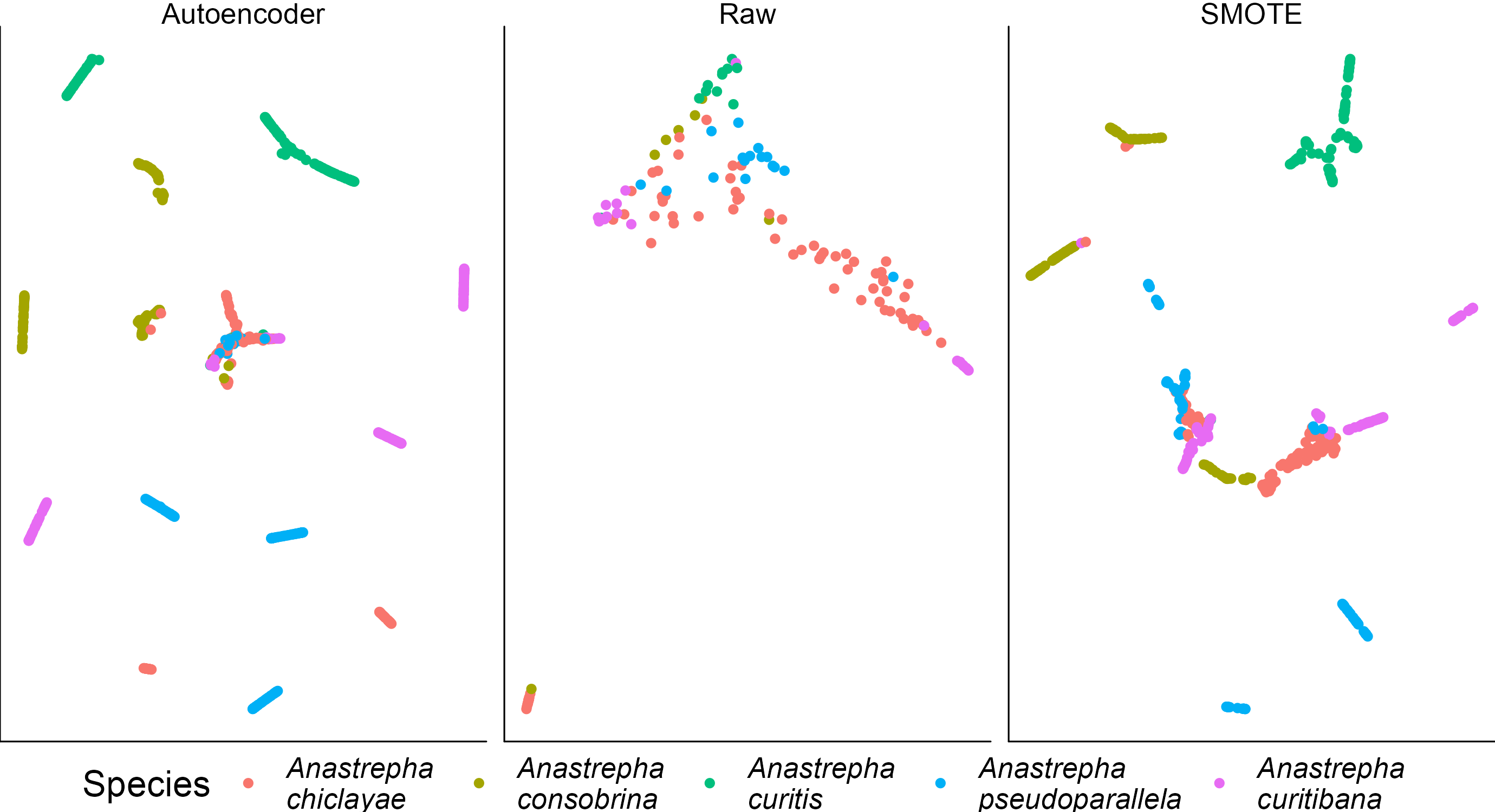}
    \caption{PaCMAP visualisation applied to the augmented TSP-based features applied to the coloured images. The colour represents each species of the \textit{pseudoparallela} flies presented in the dataset.}
    \label{TSPAgumentedFeaturePaCMAP}
\end{figure}

Figure~\ref{VGG16AgumentedFeaturePaCMAP} and Figure~\ref{TSPAgumentedFeaturePaCMAP} present the lower dimensional space estimated by the PaCMAP algorithm, showing the differences among the augmented datasets generated by the autoencoder, and SMOTE algorithms for the TSP-based and VGG16-PCA features that were obtained for coloured, gray-scale, and histogram equalised images. The PaCMAP algorithm illustrates the estimation of the distances among the data points of the higher dimension space of the collected features in solely two dimensions. The PaCMAP indicates that the feature space related to the raw dataset contains points with higher overlapping than the augmented datasets. Moreover, the autoencoder and SMOTE algorithm proved a feature space with clearer clusters, indicating a potential increase in performance for these datasets.

\section{Discussion}
\label{discussion}
We proposed a new set of features based on the shortest Hamiltonian Cycle extracted in the wing's structure and compare them with features obtained by the VGG16-PCA architecture to classify species of the \textit{A. pseudoparallela} group with various learning algorithms and data augmentation techniques. Our results showed that using the Random Forests algorithm trained with the proposed features collected from the coloured images and augmented using the SMOTE algorithm yielded higher average performance than the other presented combinations. Our results agreed with the broader literature on ensemble learning applied to imbalanced data~\citep{rezvani2023broad, khan2024review}. Our results provided a novel contribution to the \textit{pseudoparallela} group by introducing the proposed approach to classify closely related species given the challenges of imbalanced and low data availability. 

Moreover, our results indicated that the use of the proposed features collected by the solving TSP based on the wing's nodes and collecting specific distances based on the resulting shortest HC provide higher average performance compared to the use of the VGG16-PCA applied to the high-resolution images of the wings. Therefore, it presented a promising alternative for researchers looking for data dimensionality reduction and classification of winged insect species. Moreover, using coloured images instead of image transformations (gray scale and histogram equalised) showed better overall performance, indicating no need for modifying the RGB structure of the image even though for the human eyes, the histogram equalisation would emphasise the node structure of the wings. Finally, the SMOTE algorithm also increases the algorithm's overall performance in classifying this group. The PaCMaP algorithm highlighted modification on the feature space by the augmented and pre-processing methods used in the dataset~\citep{palma2023machine}. The augmented data presents clearer clusters compared to the raw dataset, which highlights the improved performances of the learning algorithms when using augmented data with both strategies. 

\textit{A. consobrina} specimens presented the most challenge for the learning algorithms, even for the best approach proposed in this paper. The lack of available images can explain this, and the increase in average performance by the SMOTE algorithm highlighted this possible explanation. Also, the classification of the studied group showed to be more challenging compared to other species of \textit{Anastrepha}, such as \textit{A. fraterculus} (Wiedemann), \textit{A. obliqua} (Macquart), and \textit{A. sororcula} Zucchi~\citep{Perre2016, leonardo2018deep} and our work highlights the importance of proposing new techniques for automatically identifying fruit fly species that face the challenges presented in this paper. Further work in obtaining additional pictures of this rare species could increase the performance of the learning algorithms. Also, we aim to incorporate data augmentation at the image level by introducing new deep generative techniques, such as Generative Adversarial Networks (GANs) and Variational Autoencoders (VARs).

\section{Conclusion}
\label{conclusions}
In this paper, we proposed a new approach to classify species of the \textit{Anastrepha pseudoparallela} group using classical computer science techniques with machine learning and computer vision methods. The main challenges faced by our work relate to the group's rare species characteristics, which results in imbalanced scarce data. We showed that combining the proposed features collected from the wing's node structure and the RGB information from specific polygons formed by spe\-cia\-lists can better classify those species than the traditional VGG16-PCA architecture based on the various learning algorithms tested. Moreover, we showed that the use of coloured images with SMOTE and Random Forests algorithms provided a higher classification performance based on these species. This study serves as an initial exploration of the classification of rare \textit{Anastrepha} species and can serve as a basis for classifying fruit fly species of other groups that present scarce data availability and class imbalance. 

\section{Acknowledgments}

This publication has emanated from research conducted with the financial support of Science Foundation Ireland under Grant 18/CRT/6049. The opinions, findings, conclusions, and recommendations expressed in this material are those of the authors and do not necessarily reflect the views of Science Foundation Ireland.

\section{Declarations}
~~~~
\textbf{Ethical Approval} Not applicable.

\textbf{Competing interests} Not applicable.

\textbf{Authors’ contributions} All authors conceived and designed the research.  

\textbf{Funding} This publication has emanated from research conducted with the financial support of Taighde Éireann – Research Ireland under Grant number 18/CRT/6049.

\textbf{Availability of data and materials} All datasets and scripts are made available at \url{https://github.com/GabrielRPalma/AnastrephaPsedoparallelaClassification}

\bibliographystyle{apalike}
\bibliography{ref}

\end{document}